\documentclass[manuscript]{aastex631}

\usepackage[utf8]{inputenc}
\usepackage{graphicx}
\usepackage{subfigure}
\usepackage{subfigure}
\usepackage{graphicx}
\usepackage{booktabs} 
\let\tablenum\relax 
\usepackage{siunitx}

\begin{document}

\title{Cocoon-to-Butterfly Transformation in Protoplanetary Nebula: Detection of a Spherical Halo and a Barrel-shaped Torus in
IRAS~06530$-$0213}

\correspondingauthor{Yong Zhang}
\email{zhangyong5@mail.sysu.edu.cn}

\author[0009-0007-4663-2643]{Hao-Min Sun}
\affiliation{School of Physics and Astronomy, Sun Yat-sen University, 2 Daxue Road, Tangjia, Zhuhai, Guangdong Province,  China}

\author[0000-0002-1086-7922]{Yong Zhang}
\affiliation{School of Physics and Astronomy, Sun Yat-sen University, 2 Daxue Road, Tangjia, Zhuhai, Guangdong Province,  China}
\affiliation{CSST Science Center for the Guangdong-Hongkong-Macau Greater Bay Area, Sun Yat-Sen University, Guangdong Province, China}

 \author{Sheng-Li Qin}
 \affiliation{School of physics and astronomy, Yunnan University, Kunming, Yunnan Province, China}

\begin{abstract}
Protoplanetary nebulae (PPNs) represent a critical evolutionary bridging stage between the asymptotic giant branch (AGB) and planetary nebula (PN) stages. Their dynamical structures provide key insights into late stellar evolution. Here, we report CO and $^{13}$CO imaging observations of the carbon-rich PPN {IRAS~06530$-$0213}, a source exhibiting the unidentified
\SI{21}{\micro\meter} emission band, conducted with the Northern Extended Millimeter Array (NOEMA). The CO maps reveal a spherical halo (diameter $\sim10''$) surrounding a central barrel-shaped torus, where the torus displays an inner diameter of $\sim1.5''$ and an outer diameter of $\sim2''$. Through three-dimensional morpho-kinematic modeling with the SHAPE software, we determine that {IRAS~06530$-$0213} experienced its final thermal pulse during the AGB phase $\sim6500$ years ago, transitioning into the PPN phase $\sim4500$ years ago. 
Our analysis indicates that the blue-shifted CO emission, a feature also detected in several other PPNs and PNs previously misattributed to the interstellar clouds, actually originates from the obscuration of the central nebula by the remnant AGB halo.
These findings are expected to deepen our understanding of the dynamical structures of PPNs, as well as their pivotal transitional role in the late-stage evolution of low-to-intermediate mass stars.
\end{abstract}


\keywords{Circumstellar matter (241) --- Protoplanetary nebulae (1301) --- Radio interferometers (1345) --- Stellar mass loss (1613) }

\section{Introduction} \label{sec:intro}

Stars with initial masses ranging from approximately 0.8 to 8 M$_{\odot}$ evolve into the Asymptotic Giant Branch (AGB) phase during their later-stage evolution. 
In this phase, the star is subjected to intense stellar winds that gradually strip its outer material. The mass loss rates during this process range 
from $10^{-8}$ to $10^{-4}$ M$_{\odot}$ yr$^{-1}$.
As a result of this continuous stripping, a circumstellar envelope forms around the central star \citep{1993ApJ...413..641V,1995A&A...299..755B,2019NatAs...3..408D}. 
As the central star's temperature subsequently rises gradually, the surrounding envelope becomes ionized, and is ultimately transformed into a planetary nebula (PN) with a hot central core. The transition from AGB to PN involves an intermediate protoplanetary nebula (PPN) phase, which persists for approximately $10^{3}$ years. 
In this phase, compared to the AGB stage, the circumstellar envelope exhibits a diverse range of asymmetric morphologies, such as bipolar, spiral, or disk-like structures \citep{1998AJ....116.1357S,2020Sci...369.1497D}. The exact mechanisms driving the diverse morphological variations of PPN circumstellar envelopes remain unclear.
Binary systems may play a crucial role in sculpting PPN morphologies
\citep[see][for a review]{2018Galax...6...99L}. 
Ample evidence indicates that PPNs display certain morphological correlations with PNs \citep{,2000ApJ...528..861U,2007AJ....134.2200S}. Leveraging this morphological link, high-resolution mapping observations of PPNs can profoundly reveal their dynamical and chemical processes, thereby offering critical insights into the key mechanisms governing the morphological evolution of PNs and the overall late-stage stellar evolution.

IRAS~06530$-$0213 was first identified as a carbon-rich PPN by \citet{2003ApJ...590.1049H} based on its spectral energy distribution (SED), spectral features, and elemental abundances. Subsequent studies by \citet{2004A&A...417..269R} and \citet{2009ApJ...694.1147H} further revealed its s-process enrichment and classified it as a typical 21\,$\mu$m source\footnote{The term ``21\,$\mu$m sources'' refer to a subclass of carbon-rich PPNs characterized by a prominent unidentified emission feature near the 20.1\,$\mu$m infrared band \citep{1989ApJ...345L..51K}.}. Recent research by \citet{2025A&A...696A.102S} has revealed a potential association between the 21\,$\mu$m feature and the presence of an equatorial density enhancement (EDE) structure in carbon-rich PPNs. Notably, high-resolution imaging observations have detected EDE structures in all four 21\,$\mu$m sources for which such data are available \citep{2000ApJ...528..861U,2009ApJ...692..402N,2012ApJ...759...61N,2025A&A...696A.102S}. However, the remaining 21\,$\mu$m objects lack comparable high-resolution observations. This observational gap impedes our capacity to assess the true prevalence of EDE structures among 21\,$\mu$m PPNs, underscoring the necessity of performing high-resolution imaging observations on additional 21\,$\mu$m sources. Beyond this specific context, such observations also provide critical insights into the dynamical and morphological evolution of late-stage stars.

In this work, we present interferometric imaging observations of IRAS~06530$-$0213. Details of the observations and data reduction are provided in Sect.\ref{sec:obser}. The spectra and images of the identified molecules, along with the morpho-kinematic model developed to characterize the nebular structure, are presented in Sect.\ref{sec:results}. In Sect.\ref{sec:dis}, we discuss the dynamical evolution of the nebula
and the influence of nebular structure on spectral line profiles. Our key conclusions are summarized in Sect.\ref{sec:concl}.

\section{Observations and data reduction} \label{sec:obser}

\begin{figure*}[h]
\centering
\includegraphics[width=0.8\linewidth]{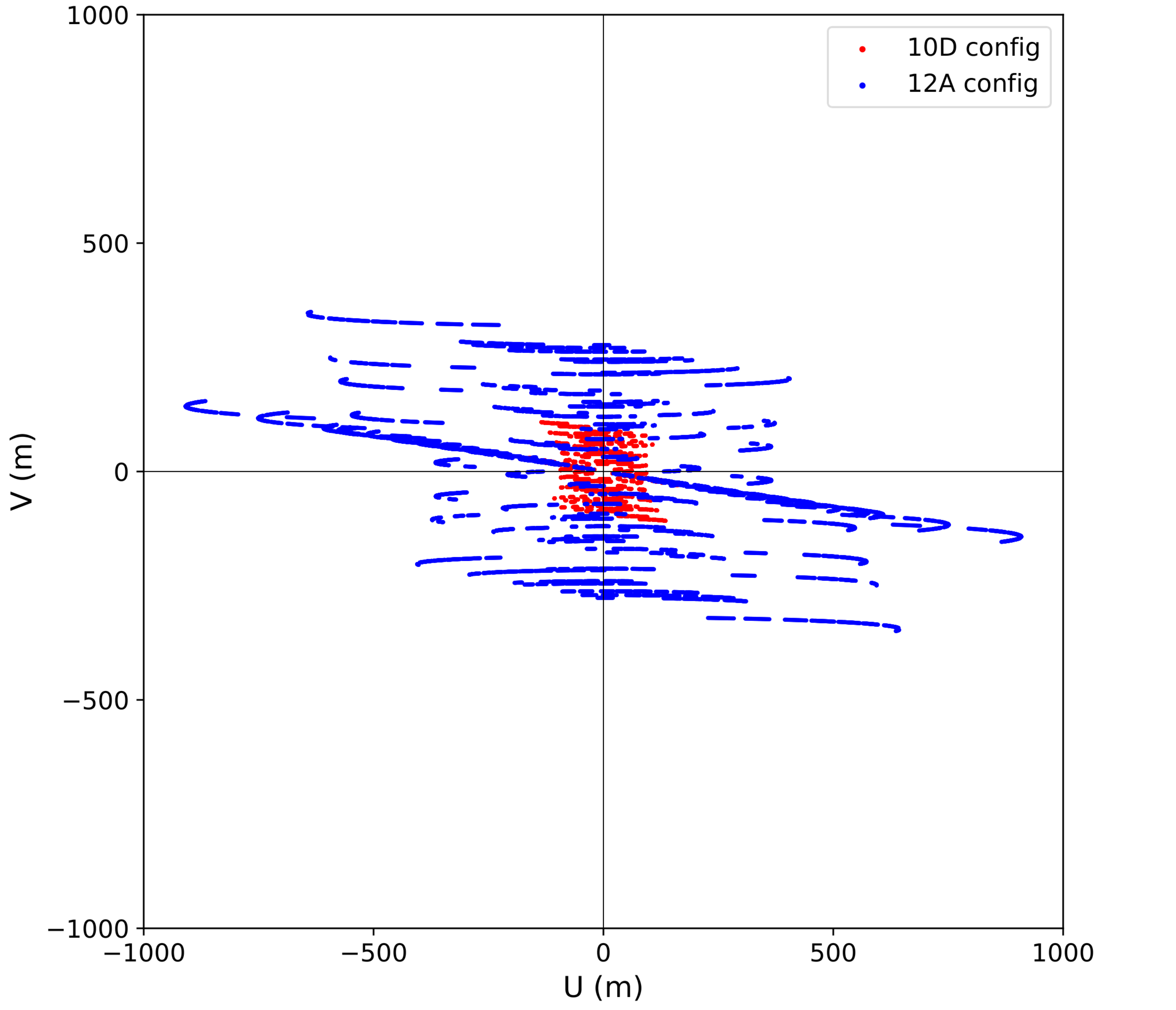}
\caption{UV coverage acquired by combining the 12A and 10D configurations.}
\label{uv}
\end{figure*}

IRAS~06530$-$0213 was observed using the Northern Extended Millimeter Array (NOEMA) under configurations 12A (project code W21BI) and 10D (project code W20BI), respectively.
The observation log is listed in Table \ref{tab1}.
The 12A configuration observations covered the Lower Sideband (LSB: 213--220 GHz) and Upper Sideband (USB: 229--236 GHz), while the 10D configuration spanned LSB: 213--222 GHz and USB: 229--237 GHz. To simultaneously capture both extended large-scale emission and high-resolution details of the central region, we combined data from the two antenna configurations: 10D (baseline lengths: 24--176 m) and 12A (baseline lengths: 32--920 m). Given discrepancies in frequency coverage and channel allocation between the datasets, we first discarded edge channels and selected common frequency ranges (LSB: 214--220 GHz; USB: 229.5--236 GHz). Identical calibration procedures were then applied to ensure consistent channel number, bandwidth, and central frequency. After weighting normalization, the data were resampled and merged. The resulting UV coverage is shown in Figure \ref{uv}. For the combined dataset, Doppler tracking established a central velocity of $33\, \rm km\,s^{-1}$ relative to the Local Standard of Rest (LSR) (kinematic definition), with the phase center set at $\left(\alpha_{\rm J2000},\delta_{\rm J2000}\right) = \left(06^{\rm h}55^{\rm m}31\fs820,\, -02^\circ17'28\farcs300\right)$.
 
All data were calibrated using the \texttt{GILDAS/CLIC} software suite\footnote{\url{http://www.iram.fr/IRAMFR/GILDAS}}. Calibration sources are listed in Table \ref{tab1}. Imaging and deconvolution were performed with \texttt{GILDAS/MAPPING} and \texttt{GILDAS/IMAGER}\footnote{\url{https://imager.oasu.u-bordeaux.fr}}, employing the CLEAN algorithm of \citet{1974A&AS...15..417H}. Uniform weighting was adopted to optimize spatial resolution. The synthesized beams for the 10D configuration, 12A configuration, and combined spectral line data cubes are $1.91'' \times 1.29''$, $0.67'' \times 0.3''$, and $0.72'' \times 0.41''$, respectively. The spectral resolution for all observations is approximately 2 MHz ($2.7\, \rm km\,s^{-1}$) at 1.3 mm (NOEMA Band 3).

\section{Results} \label{sec:results}

\subsection{Spectral lines}\label{speclines}

Previously reported single-dish observations targeting  IRAS~06530$-$0213 have identified emission lines of CO ($J=2$--1 and $J=1$--0), $^{13}$CO ($J=2$--1), HCN ($J=3$--2), HNC ($J=3$--2), and CN ($N=2$--1) \citep{2005ApJ...624..331H,2024AJ....167...91Q}. Leveraging the enhanced angular resolution and sensitivity of the NOEMA interferometric observations, we have now detected ten distinct emission features belonging to eight molecular species and isotopologues. Notably, C$^{18}$O, HC$_3$N, C$_4$H, SiC$_2$, SiS, and CH$_3$CN are discovered for the first time in this source, significantly expanding the inventory of molecular tracers available for studying its circumstellar environment. 
 
Given the primary objective of this work,
investigating the kinematic structure of this 
21\,$\mu$m PPN, we focus primarily on the results for CO and $^{13}$CO, which are well-established tracers of large-scale nebular dynamics. The spectral characteristics and high-resolution imaging of the remaining species, which are crucial for constraining nebular chemistry, will be presented in Sun et al. (in preparation), a work dedicated to the chemical composition of the circumstellar medium.
 
\begin{figure*}
\centering
\includegraphics[width=0.9\linewidth]{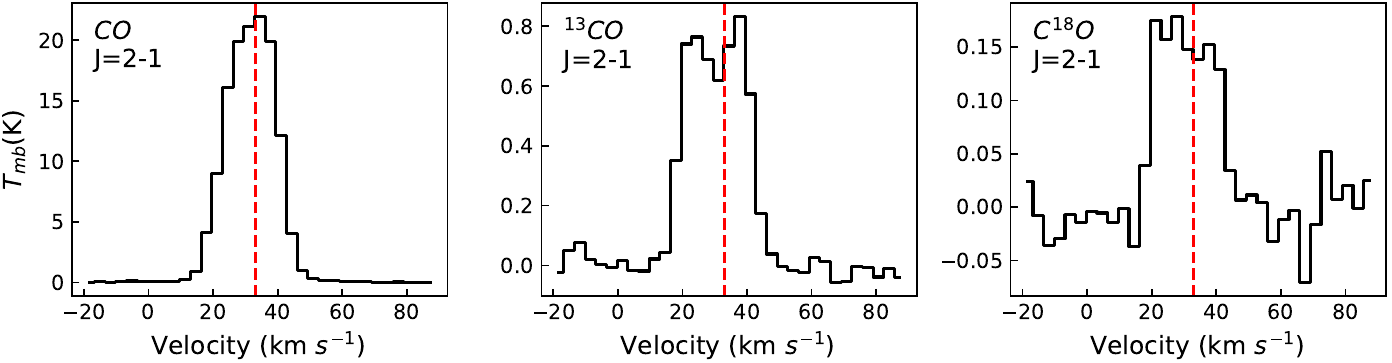}
\caption{CO ($J=2$--1) line profile and its isotopologue line profiles detected in IRAS~06530$-$0213, with 
 the systemic velocity ($33$\,km~s$^{-1}$) inidicated by a vertical dashed line.}
\label{co_spec}
\end{figure*}

Figure~\ref{co_spec} illustrates the line profiles of CO and its isotopologues, derived by averaging flux densities over a $2\arcsec\times2\arcsec$ aperture centered on the phase center from the combined data. The spectral line measurements are presented in Table~\ref{line}, where E$_u$ is the energy of the upper level, and  $I_{\rm peak}$ and $\int I \, dv$ represent the peak and integrated intensities, respectively. The $^{12}$CO ($J=2$--1) line exhibits an optically thick, centrally saturated parabolic profile. In contrast, the $^{13}$CO and C$^{18}$O lines display optically thin profiles, reflecting their reduced opacity relative to $^{12}$CO. These contrasting opacity properties imply that while $^{12}$CO ($J=2$--1) is effective for tracing the extended, low-density outer regions of the nebula, its saturation at the line center limits its utility for probing the inner, high-density structural components. \citet{2024AJ....167...91Q} previously detected a narrow  feature (2\,km~s$^{-1}$ in width) at  $\sim20$\,km~s$^{-1}$ in single‐dish CO ($J=1$--0) observations, which they attributed to
the influence of a nearby interstellar cloud\footnote{Please note that this feature lies on the blue side of the line at $\sim$ 20\,km~s$^{-1}$. It should not be confused with the strong line at $\sim$ 28\,km~s$^{-1}$.}. In our NOEMA CO ($J=2$--1) spectra, this feature is manifested as a slight asymmetry in the line profile. The origin of this feature will be discussed in Section~\ref{co}.

\subsection{Maps} \label{maps}

\begin{figure*}
\centering
\includegraphics[width=0.8\linewidth]{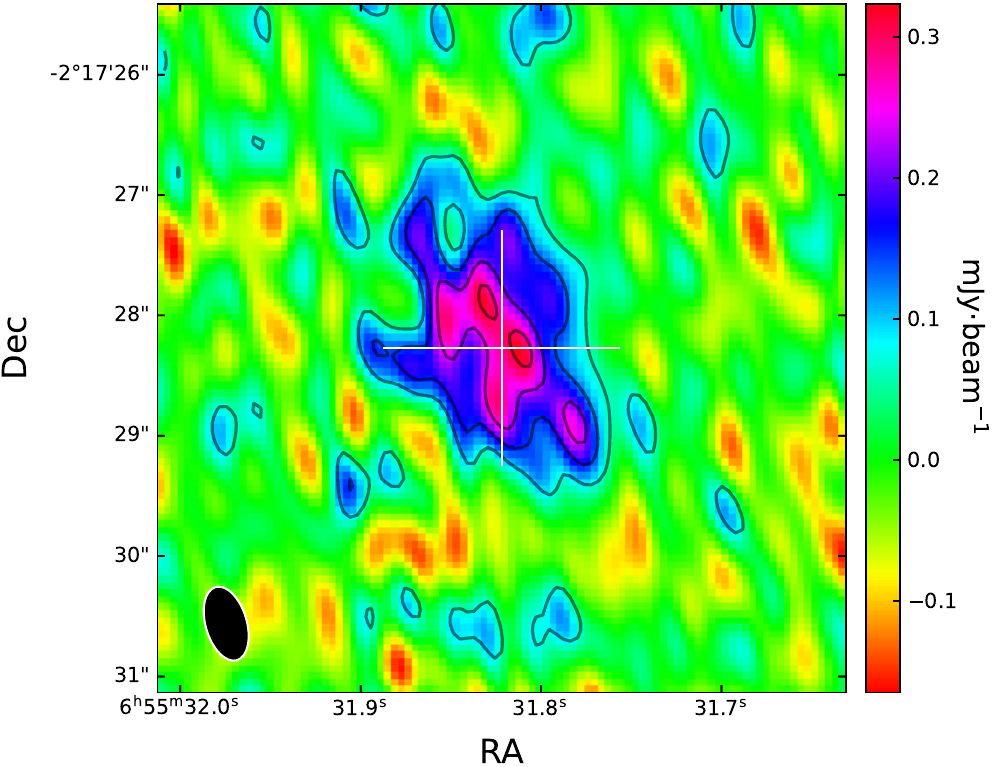}
\caption{Continuum emission at 1.3\,mm wavelengths. The synthesized beams shown in the lower-left corner have a size of $0.7\arcsec \times 0.4\arcsec$. The cross denotes the position of the phase center.
Contour levels are set at $1\sigma$ intervals starting from $1\sigma$, where the rms $\sigma = 62.5\,\mu\rm{Jy\,beam^{-1}}$.
}
\label{cont}
\end{figure*}

\begin{figure*}
\centering
\includegraphics[width=0.8\linewidth]{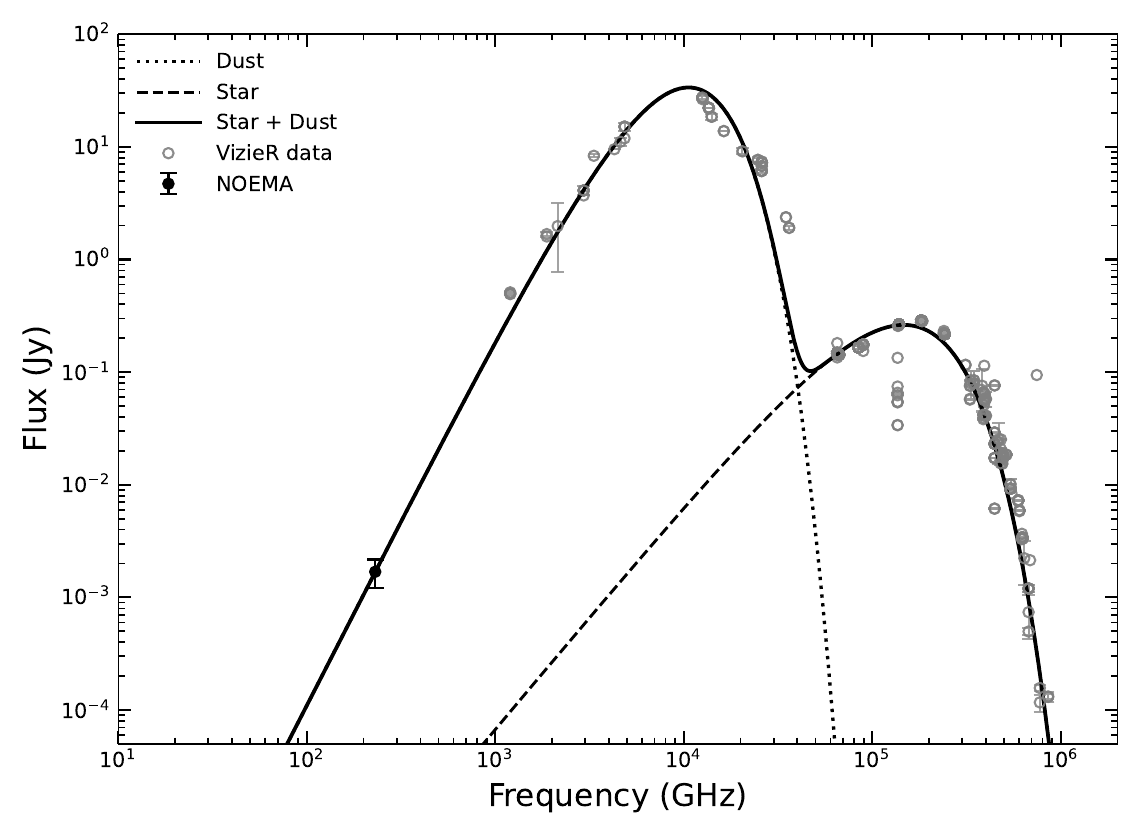}
\caption{SED and two-component fitting of IRAS 06530$-$0213.
}
\label{sed}
\end{figure*}

Figure~\ref{cont} displays the continuum emission map derived from the combined multi-channel data at 1.3 mm wavelength. Owing to its low signal-to-noise ratio, the continuum image exhibits a lack of clearly resolved structural features.
To estimate the flux density of the continuum emission, we constrained the frequency range to two intervals (213.5--217.5 GHz and 229--236 GHz), from which CO lines were masked out. This yielded a total integrated flux density of 1.69 mJy.
The uncertainty introduced by flux calibration ($\sigma_{\rm c}$) is $\sim10\%$, corresponding to $\sigma_{\rm c} = 0.17$ mJy, while the measurement error is $\sigma_{\rm m} = 0.45$ mJy. The total uncertainty in the continuum emission flux density is therefore 
$\sigma_{\rm tot} = \sqrt{\sigma_{\rm m}^2 + \sigma_{\rm c}^2} = 0.48 \, \text{mJy}$,
indicating that the error budget is dominated by noise rather than the  flux calibration. We constructed the SED by integrating multi-wavelength photometric data from the VizieR database\footnote{\url{https://vizier.cds.unistra.fr/}} (see Figure~\ref{sed}). It exhibits a clear double-peak structure that aligns well with the radiative mechanisms of the central star and circumstellar dust. The high-frequency peak corresponds to the central star’s thermal radiation, while the low-frequency peak arises from the thermal re-radiation of circumstellar dust. We applied a two-component model to fit this SED: the hot component associated with the central star was fitted using a Planck function, which yielded an  temperature of approximately 2500 K, and the cold component linked to circumstellar dust was fitted with a modified Planck function, which resulted in an average circumstellar dust temperature of around 120 K. Notably, our NOEMA measurement falls precisely within the tail of the cold component, and this alignment leads to the definitive conclusion that the continuum emission from our NOEMA observations originates from the thermal emission of circumstellar dust. Additionally, SED fitting yields a dust emission spectral index of 3.3, which corresponds to a dust emissivity index of $\beta = 1.3$.
This emissivity index is lower than the value typically adopted for interstellar dust 
\citep[$\beta \sim 2$, e.g.,][]{1983QJRAS..24..267H, 2007ApJ...657..810D, 2014A&A...571A..11P}. 
One possible explanation is that the dust grains surrounding this source may be larger than typical interstellar dust grains \citep{2006ApJ...636.1114D, 2014prpl.conf.....B}, with this size difference probably driven by dust growth processes in the circumstellar environment.

\begin{figure*}
    \centering
    \begin{minipage}{0.85\linewidth}
        \centering
        \includegraphics[width=\linewidth]{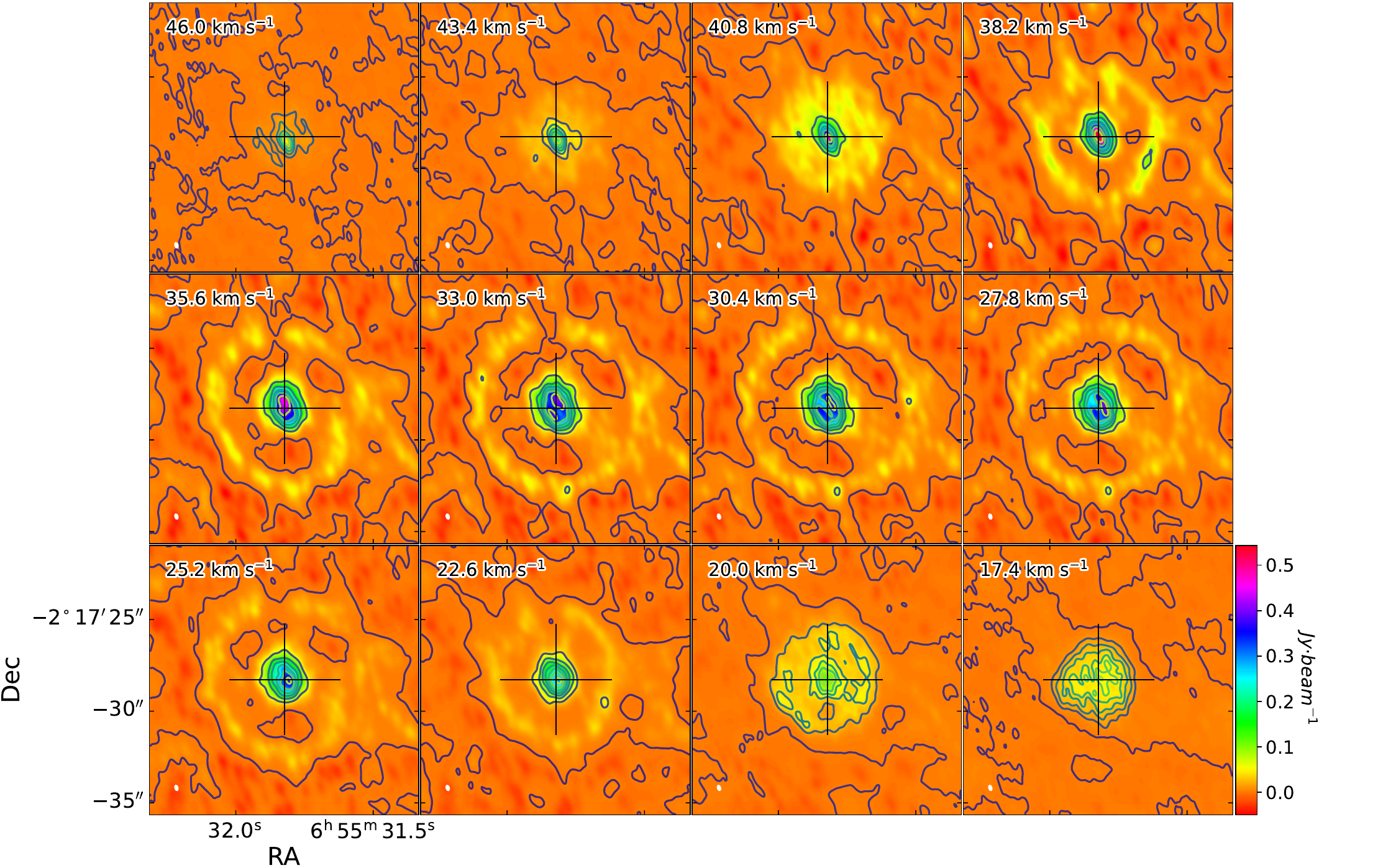}
    \end{minipage}
    \hfill
    \begin{minipage}{0.85\linewidth}
        \centering
        \includegraphics[width=\linewidth]{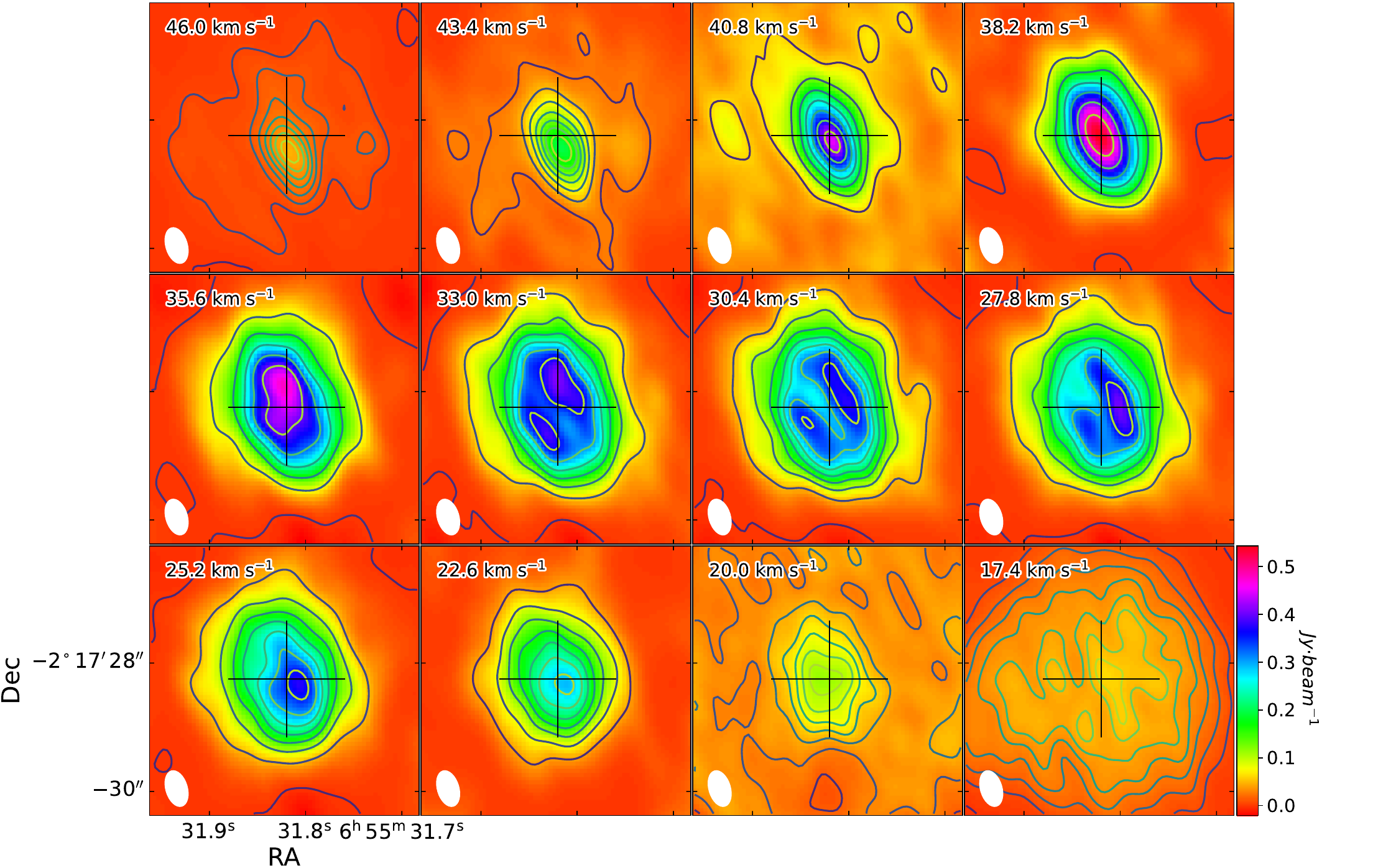}
    \end{minipage}
    \caption{Channel maps of CO ($J=2$--$1$). The upper panels cover $14.5\arcsec \times 14.5\arcsec$, while the lower panels present a zoomed-in view spanning $4.1\arcsec \times 4.1\arcsec$. Contour levels are individually defined for each channel by ten equally spaced lines between the channel's intensity minimum and maximum. Crosses mark the phase center position, and LSR velocities are labeled in each panel. The object's systemic velocity is $33\,\rm{km~s^{-1}}$. The white ellipse in the bottom-left corner denotes the synthesized beam, with dimensions $0.72\arcsec \times 0.41\arcsec$ and a position angle (PA) of $15^\circ$.
 }
    \label{co_channel}
\end{figure*}

Figure~\ref{co_channel} presents the continuum-subtracted channel maps of CO ($J=2$-1) from the combined 12A and 10D observations. Examination of the upper panel in Figure~\ref{co_channel} reveals that the nebula comprises a large-scale halo encircling a smaller, compact central structure. The halo does not exhibit perfect symmetry; irregular emission in its southwest corner gives it an overall Q-shape. This halo appears across channels spanning 
22.6--38.2\,km~s$^{-1}$, attains its maximum extent ($\sim10\arcsec$) at the systemic velocity (33\,km~s$^{-1}$), and progressively contracts with increasing velocity. Analogous halo structures in PPNs, attributed to transient, super-high mass-loss episodes during the late AGB phase, have been documented in multiple sources thourgh Hubble Space Telescope ({\it HST})  imaging surveys of PPNs \citep{2007AJ....134.2200S}, such as IRAS~17253$-$2831, IRAS~18276$-$1431, and IRAS 23304+6147.
The halo appears relatively thin and shows clumpy, fragmented features that might reflect hydrodynamic instabilities at the interface between the fast stellar wind and the surrounding gas. In such a scenario, shock‐driven turbulence arising from the wind colliding with slower‐moving circumstellar material could give rise to Rayleigh–Taylor or Richtmyer–Meshkov instabilities, potentially breaking the halo into dense clumps. Such instabilities are typical signatures of mass-loss dynamics during the AGB-PPN transition, where stellar wind interactions shape the nebular morphology on timescales of thousands of years.
Alternative explanations include the possibility that the ejecta expelled from the AGB star are inherently clumpy. 

\begin{figure*}
\centering
\includegraphics[width=1\linewidth]{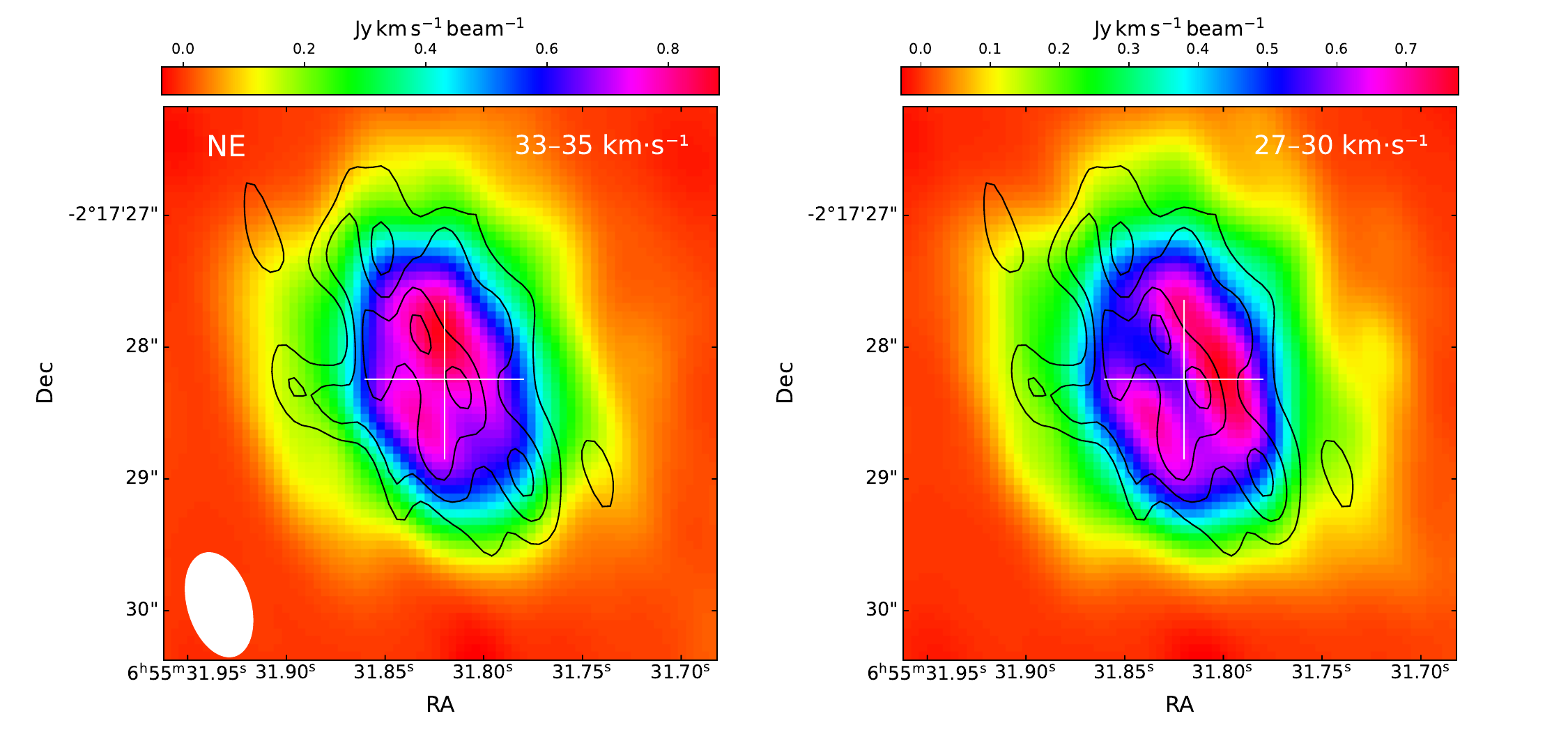}
\caption{Integrated intensity maps of the CO ($J=2$--1) line over two different velocity ranges. The northeast (NE) direction 
    is marked in the upper-left corner of the map. The black contours represent the continuum emission, as illustrated in Figure~\ref{cont}. The white ellipse at the bottom-left corner represents the synthesized beam, with a size of $0.72\arcsec \times 0.41\arcsec$  and a PA of $15^\circ$.}
\label{torus_map}
\end{figure*}

The lower panel of Figure~\ref{co_channel} focuses on the nebula's compact central structure, which manifests as an inclined torus (outer radius $\sim1.5\arcsec$) across the velocity range 27--35\,km~s$^{-1}$. At blueshifted and redshifted velocities, an ellipsoidal envelope (EE) encircling the torus becomes apparent. The torus, elongated along the northeast-southwest axis in sky-plane projection, corresponds to the limb-brightened bipolar lobes previously identified via {\it HST} observations of this source \citep{2000ApJ...528..861U}. 
Figure~\ref{torus_map} shows integrated intensity maps of the torus across two distinct velocity intervals. The results reveal that the southwest segment of the torus vanishes at the red velocity end (Figure~\ref{torus_map}, left panel), while the northeast segment is absent at the blue velocity end (Figure~\ref{torus_map}, right panel). These observations imply that the torus is viewed nearly edge-on: its pronounced edge thickness imparts a barrel-like geometry, and the barrel walls enhance the column density of CO molecules, giving rise to the observed bipolar lobe morphology. Such a barrel-like torus structure has also been  detected in infrared observations of the PN  NGC 1514 \citep{2010AJ....140.1882R,2025AJ....169..236R}, 
though its diameter-to-height ratio is significantly larger than that of IRAS~06530$-$0213. This discrepancy may arise from the evolutionary expansion of the torus along the equatorial plane.

The torus in IRAS~06530$-$0213 is  analogous to the EDE detected by \citet{2025A&A...696A.102S} in another
21\,$\mu$m PPN IRAS~23304+6147. 
Such a structure is closely tied to  binary system evolution and likely serves as a precursor to the bipolar outflows observed in evolved PPNs, mediating mass ejection and shaping the subsequent circumstellar architecture.
The carrier of the 21\,$\mu$m feature might be associated with
large carbon-bearing molecules.
The NOEMA observations of IRAS~06530$-$0213 reveal that carbon-chain species such as HC$_3$N and C$_4$H are predominantly distributed 
in the inner torus regions, in contrast to the distribution of
 CO and $^{13}$CO. Combining images traced by different molecules suggests a flared disk structure.
 This emphasizes that morphological investigations of PPNs should rely on multiple molecular tracers.
The spatial distribution, kinematics, and chemical properties of these molecules will be elaborated in a forthcoming paper (Sun et al., in preparation).

\begin{figure*}
\centering
\includegraphics[width=0.9\linewidth]{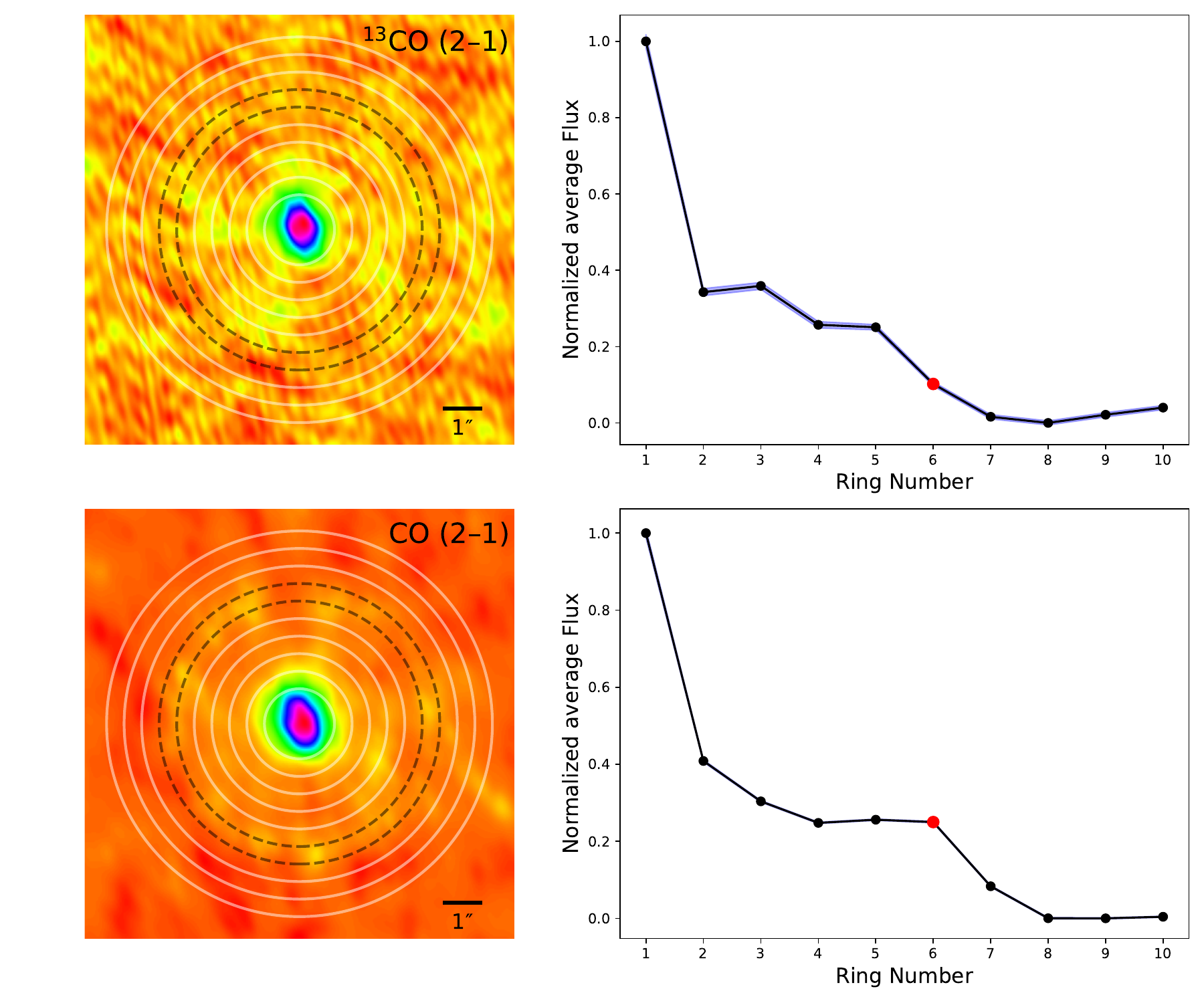}
\caption{Flux density statistics of $^{13}$CO and CO moment 0 map in different regions. \textbf{Left panels:} $^{13}$CO and CO moment 0 map. Nine concentric annular regions centered at the phase center are outlined by white solid lines and black dashed lines; of these, the black dashed ring marks the halo position at the systemic velocity. \textbf{Right panels:} Statistics of the normalized average flux within each ring from the left panel. Ring numbers 2--10 correspond to the concentric rings in the left panel, ordered from smallest to largest, with ring number 1 designating the central circular region. 
The red filled circles denote the positions of the spherical halo at the systemic velocity.}
\label{ring_13co}
\end{figure*}

\begin{figure*}
\centering
\includegraphics[width=\linewidth]{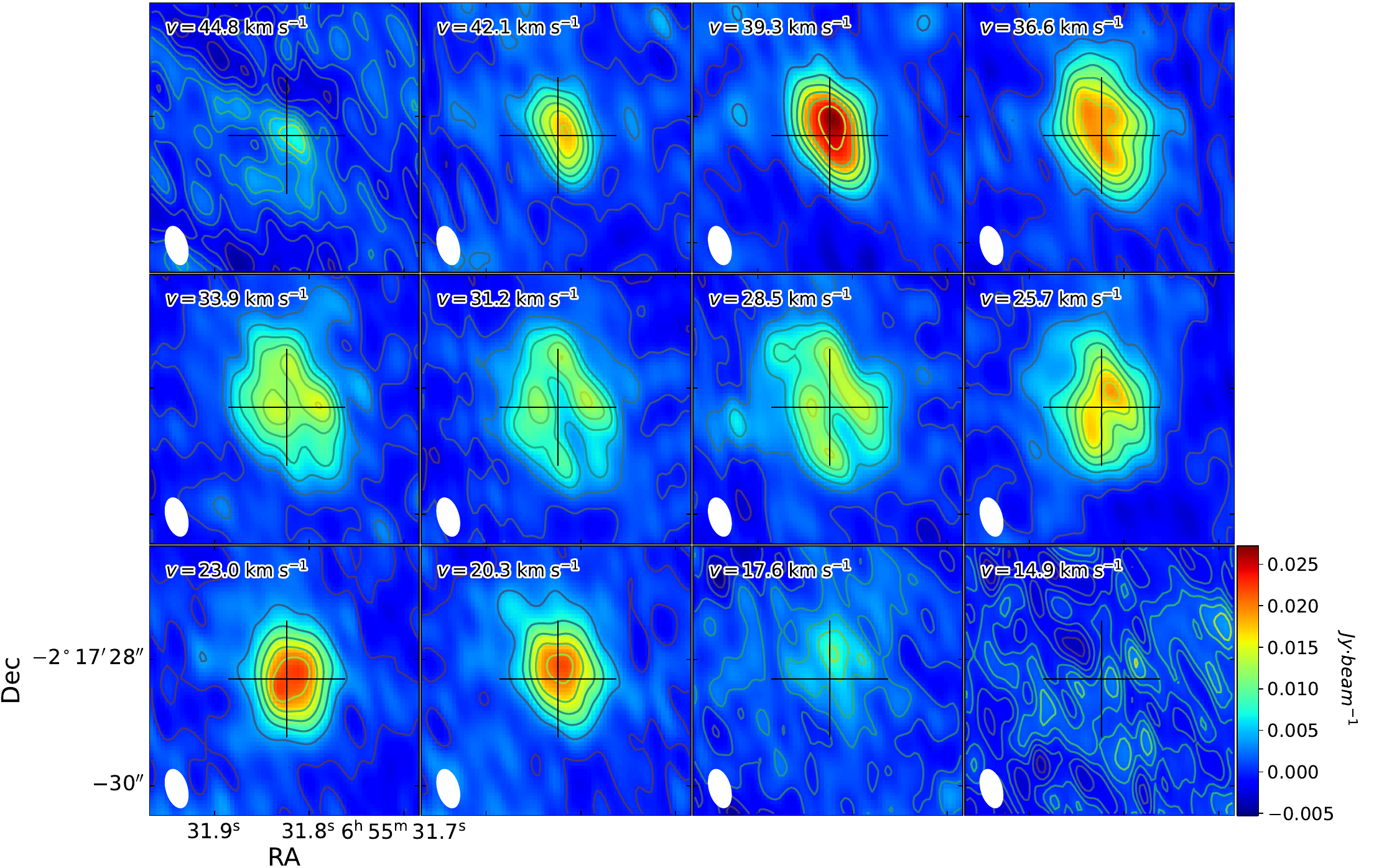}
\caption{Same as lower panels of Fig.~\ref{co_channel},
    but for $^{13}$CO $J=2-1$.}
\label{13co_channel}
\end{figure*}

No halo component analogous to that in the outer CO envelope can be seen in the $^{13}$CO ($J=2$--1) observations. To confirm this result, we constructed a $^{13}$CO moment 0 map and overlaid the region corresponding to the CO halo at the systemic velocity (marked by the black dashed circle in Figure~\ref{ring_13co}). The average flux density at this location, depicted by the red filled circle in the right panel of Figure~\ref{ring_13co}, exhibits no enhancement. We calculated the peak flux ratio of the compact central structure to the halo in CO emission as 10.2. Under the assumptions of a uniform excitation temperature and that the central CO emitting structure is optically thick, this ratio corresponds to the inverse of the halo’s optical depth. This, in turn, implies that the optical depth of CO in the halo is approximately 0.1.
If we further assume that a $^{13}$CO halo exists at the same location, yet with a peak flux at or below the rms noise level, the peak‐flux ratio between the compact central $^{13}$CO structure and its hypothetical halo must then be 
$\geq14$. Given that $^{13}$CO is optically thin, this ratio  likely serves as an indicator of the
column density contrast between the central region and the halo. Extending this contrast to the CO emission implies that the  optical depth of the compact central CO structure is $\geq1.4$. In other words, if the actual CO optical depth at the center is greater than 1.4,  even if a $^{13}$CO halo were present, its emission would be submerged in the noise and thus remain undetectable.
Nevertheless, a torus structure analogous to that in CO was detected in the $^{13}$CO envelope, as illustrated in the continuum-subtracted channel map of Figure~\ref{13co_channel}.

\subsection{Morpho-kinematic model}

\begin{figure*}
\centering
 \includegraphics[width=\linewidth]{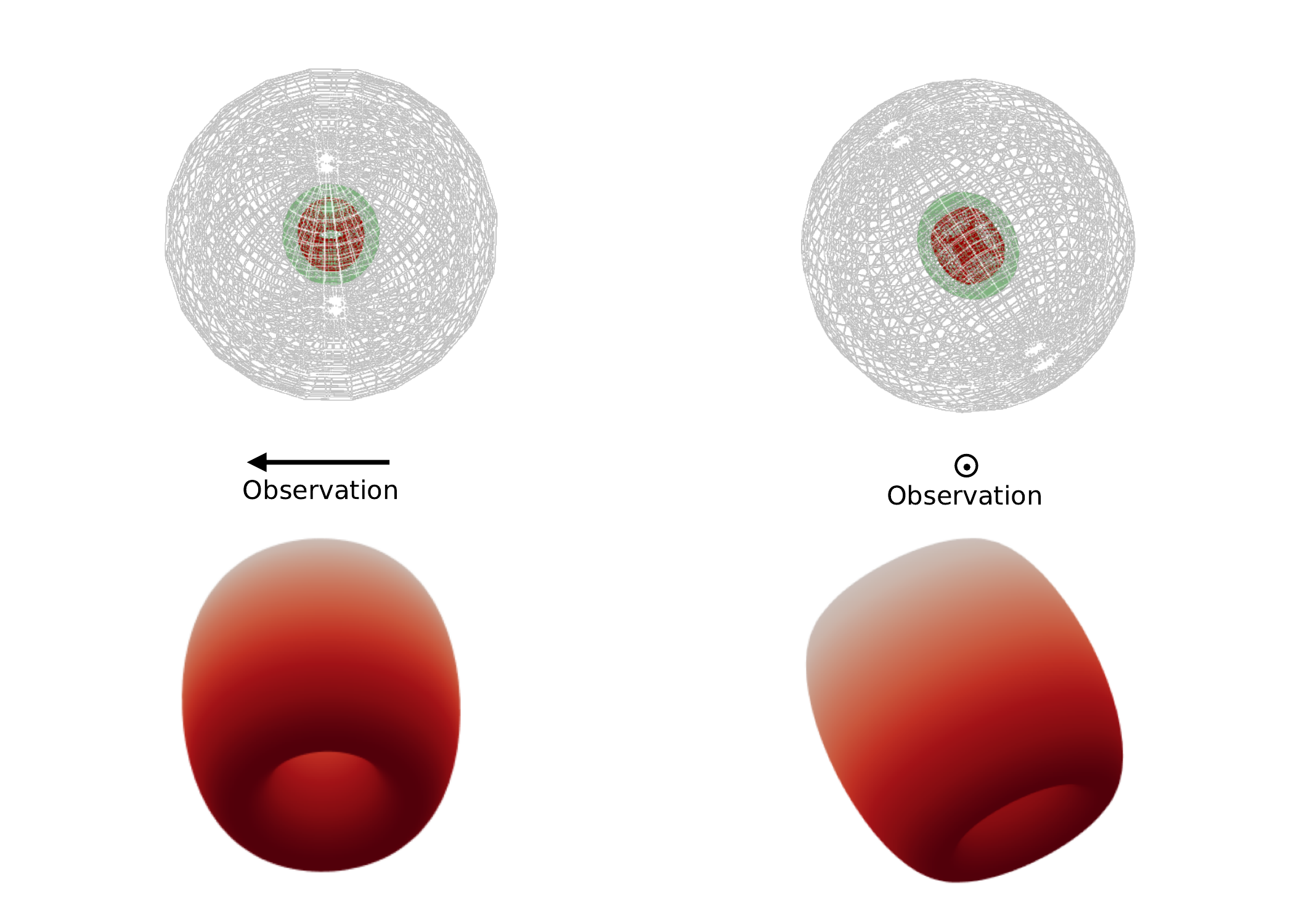}
\caption{Three-dimensional mesh geometry of the SHAPE model viewed from two orthogonal directions. 
The upper panels depict the full nebular structure: the left panel presents a side-on view from the observer's vantage point, whereas the right panel displays a front-facing view directed toward the observer. Red indicates the barrel-shaped torus, green represents the EE, and gray denotes the outer halo. The lower panels provide a close-up view of the barrel-shaped torus.}
\label{model}
\end{figure*}

To further investigate the structures of IRAS~06530$-$0213, we utilized the three-dimensional morpho-kinematic modeling code SHAPE \citep{2012ascl.soft04010S} to replicate its nebular kinematics. Specifically, we first constructed an initial nebular model with morphological features closely matching the observations, then incorporated a set of free parameters. Through iterative refinement, we ultimately generated channel maps and position–velocity (P–V) diagrams that closely replicate the observed data. This approach has been successfully applied to reconstruct PPNs in previous studies \citep[e.g.,][]{2011ApJ...740...27K,2012ApJ...759...61N,2015A&A...573A..56S,2019A&A...629A...8T,2025A&A...696A.102S}. The model comprises three components, corresponding to the torus, EE, and halo, as illustrated in Figure~\ref{model}. Following \citet{2025A&A...696A.102S}, 
we assign  both the torus and the EE their own Gaussian density profiles, along with corresponding velocity fields that adhere to Hubble law. The halo structure, confined to the envelope’s outer region and extremely thin along the line of sight, is assigned constant density and velocity values.

Given that the model is designed exclusively to reproduce the nebular kinematic structure without incorporating radiative-transfer calculations, we must map the model’s density distribution onto the molecular line brightness. This method relies on the assumption that the molecular lines are optically thin, such that surface brightness is linearly proportional to column density. Although the halo structure was detected exclusively in CO, our calculations in Section~\ref{maps} indicate that the optical depth of the halo is approximately 0.1, confirming that the CO halo emission is optically thin. Comparison of $^{13}$CO and CO channel maps reveals no significant morphological discrepancy in the compact central structure within the field of view. Furthermore, the higher signal-to-noise ratio of CO data provides a clearer tracer of nebular morphology. Consequently, we adopt CO channel maps to constrain the model's density distribution and velocity fields. This approach is validated by two factors: the outer halo emission being optically thin, and the compact central structure showing morphological consistency with $^{13}$CO, indicating minimal optical depth effects.

\begin{figure*}
\centering
\includegraphics[width=\linewidth]{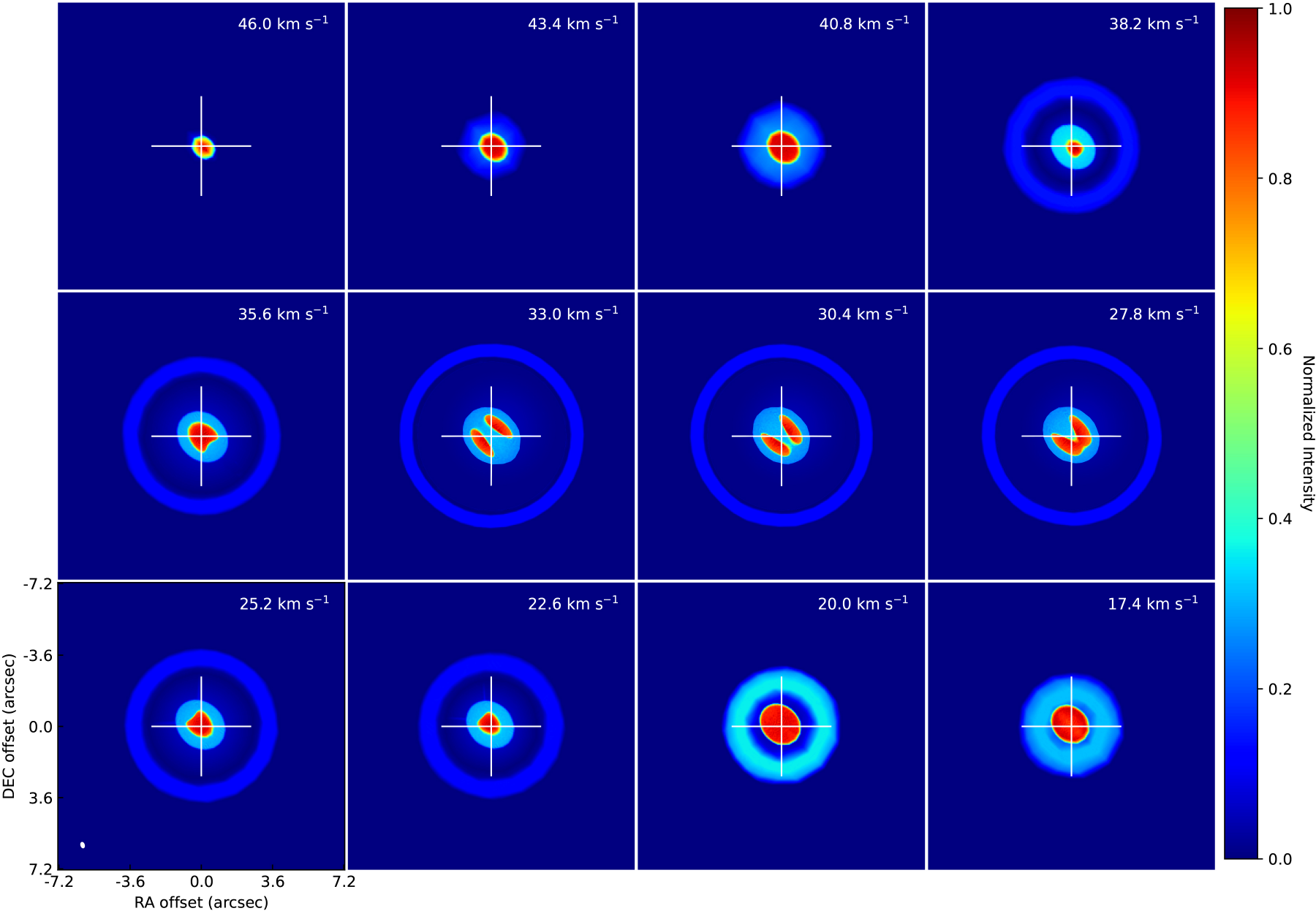}
\caption{Modeled CO ($J=2$--1) channel maps.
 The white ellipse in the lower-left corner of the image represents the synthesized beam.  The angular size of each panel is identical to
 those in the lower panels of Fig.~\ref{co_channel}.}
\label{model_channel}
\end{figure*}

\begin{figure*}
\centering
 \includegraphics[width=\linewidth]{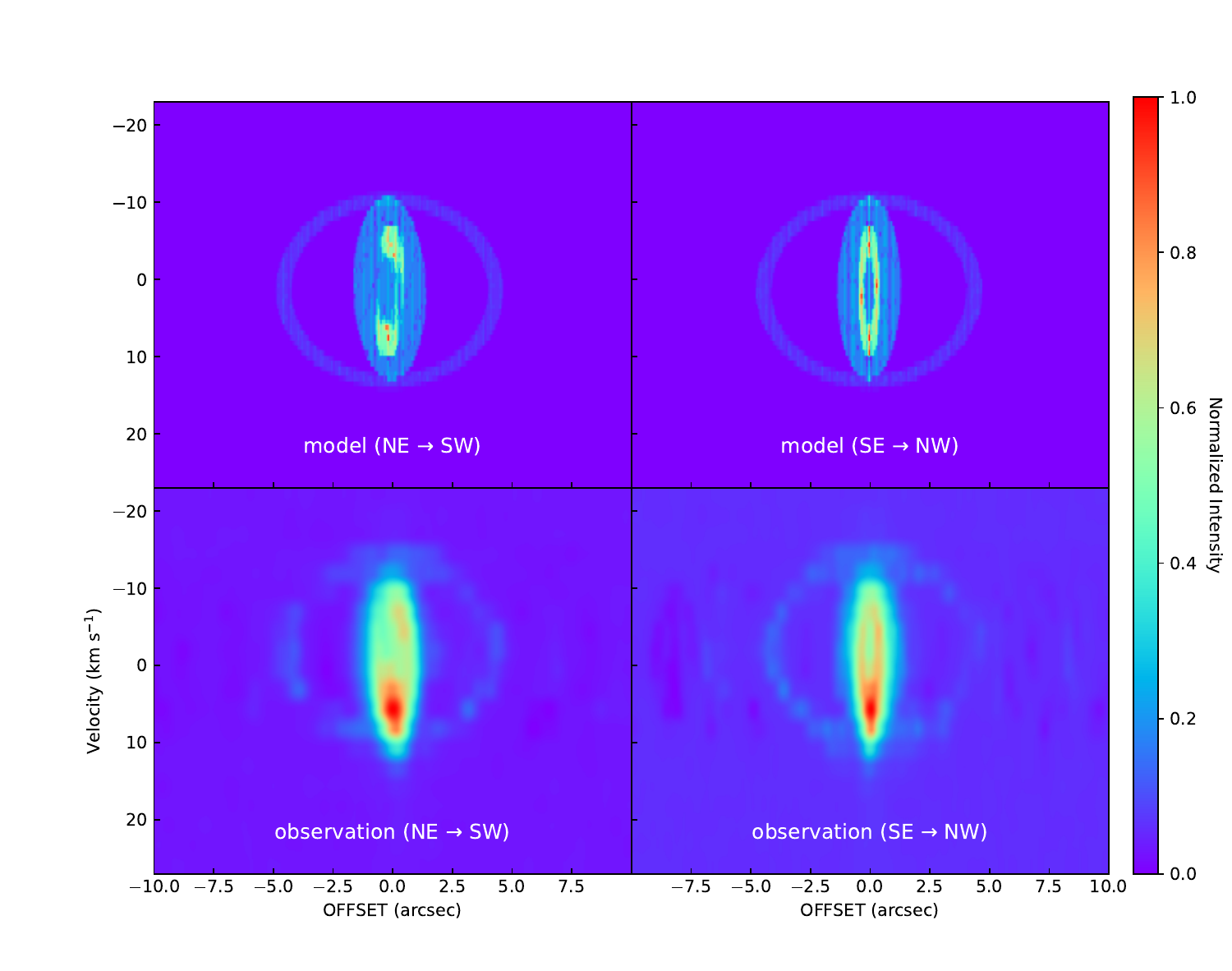}
\caption{P-V diagrams of the CO $J=2$--1 transition from model predictions (upper panels) and observations (lower panels). The left panels depict the velocity structure along the major (symmetry) axis of the torus, aligned from NE to SW, where negative offsets correspond to the NE direction. The right panels display the velocity structure along the minor axis of the torus, aligned from SE to NW, with negative offsets corresponding to the SE direction.}
\label{pv}
\end{figure*}

Figures~\ref{model_channel} shows the modeled channel maps. The model effectively reproduces the primary nebular structures observed in the channel maps, with morphological and velocity features broadly consistent with the observations, although it does not fully replicate the nebular surface brightness.
Figure~\ref{pv} shows a comparison between the observed P-V diagram and the modeling results. The kinematic structures of the three components are well traced by the P-V diagrams of CO. No signature of a rotating disk, detected in a handful of PPNs \citep{2021A&A...648A..93G}, is observed. The deduced inclination angle of the torus relative to the sky plane is approximately 20$^\circ$. The model results indicate that the maximum expansion velocities of the torus and EE are similar, at $\sim10$\,km~s$^{-1}$ and $\sim12$\,km~s$^{-1}$ respectively. For the halo structure, the velocity is approximately 15\,km~s$^{-1}$. 
At a distance of 4.1~kpc \citep{2020yCat.1350....0G}, the torus has an inner radius of 0.8\arcsec, an outer radius of 1.5\arcsec, and a maximum thickness of 2\arcsec, corresponding to 3400~au, 6300~au, and 8500~au, respectively. The EE's major and minor axes measure $\sim$ 10600~au and 9800~au, respectively. Based on the measured scales of the torus and the EE, we infer dynamical ages of $\sim$ 1500 years for the torus and $\sim$ 2000 years for the EE. The outer halo appears at a distance of $\sim40000$~au from the central star, implying a dynamical age of $\sim$ 6500 years for the halo. Its thickness of $\sim$ 2000~au might indicate that this major mass-loss event persisted for 
$\sim$ 650 years\footnote{The estimation is relatively approximate due to the complex hydrodynamic interactions between the stellar winds.}.
By comparing the dynamical ages of the EE ($\sim$ 2000 years) and halo ($\sim$ 6500 years), we infer that IRAS~06530$-$0213 initiated its transition from the AGB to the PPN phase approximately 4500 years ago.

It should be noted that the kinematic solution derived by SHAPE cannot be guaranteed to be unique.
In SHAPE, parameters such as the velocity field, density distribution, and component dimensions are all interdependent; tweaking the effect of any one parameter can be compensated by adjustments in others, still reproducing the same synthetic results. For instance, increasing the radial density gradient could be offset by modifying the expansion velocity profile, yielding identical channel map morphologies despite distinct physical configurations. This degeneracy presents challenges for constraining absolute physical properties, as multiple parameter combinations may equally well match the observed data.  

\subsection{Nebular mass}

Based on the methodologies presented by \citet{1999A&A...347..194O} and \citet{2006ApJ...645..605C}, the nebular mass can be derived from optically thin lines. For IRAS~06530$-$0213, we utilized the $^{13}\mathrm{CO}\ (J=2$--1) line to calculate the mass of the central compact structure (comprising the torus and EE). The compact CO emission at the nebular center is optically thick, so its excitation temperature can be approximated by the peak brightness temperature of $\sim 20\ \mathrm{K}$. Under Local Thermodynamic Equilibrium conditions, this value was adopted as the excitation temperature for $^{13}\mathrm{CO}$.  Assuming a $^{13}\mathrm{CO}$ fractional abundance of $f_{^{13}\mathrm{CO}} = 2 \times 10^{-5}$, we derived a mass of $1.5 \times 10^{-3}\ M_{\odot}$ for the central compact structure of IRAS~06530$-$0213. Notably, the presence of the halo necessitates its inclusion in the total nebular mass estimate. In the optically thin regime of the CO halo emission, adopting the canonical CO abundance in AGB and PPN envelopes of $f_{\mathrm{CO}} = 2\times10^{-4}$ \citep{2001A&A...377..868B} and an excitation temperature of $20\ \mathrm{K}$, we derive a halo mass of approximately $4\times10^{-3}\ M_\odot$. Therefore, the total nebular mass is estimated to be close to $5.5\times10^{-3}\ M_\odot$.

\section{Discussion}\label{sec:dis}

\subsection{Dynamical evolution}

The transition of stars from the nearly spherical symmetry of the AGB phase to the distinctly non-spherical geometries of the PN phase represents a fundamental question in late stellar evolution. The key to answering this query undoubtedly resides in the transitional PPN stage. Several PPNs have been identified that exhibit characteristics of both the AGB and PN phases; some display extreme non-spherical asymmetries in their morphological and kinematic properties \citep{2001A&A...377..868B,2007AJ....134.2200S}, while others feature ionized structures near the central star, marking the onset of the PN phase \citep{2013A&A...556A..35T,2017A&A...603A..67S,2024ASPC..536..123C}.

Our observations of IRAS~06530$-$0213 reveal it as an exemplary PPN showcasing transitional features.
The nebula’s outer halo exhibits spherical symmetry, a characteristic feature of an AGB envelope, while the inner regions display pronounced asphericity-typical of most PNs. The age difference between the halo and the central nebula 
is only $\sim 4500$ years, 
indicating that IRAS 06530$-$0213 is a relatively young post-AGB object. Its outer halo was 
likely formed during the high-mass-loss episode associated with the final thermal pulse of the AGB phase, which produced a detached shell. While the halo’s thickness 
tentatively suggests a thermal pulse duration of $\sim$ 650 years, the clumpy and gaped structure of the halo, indicative of complex hydrodynamic interactions, implies that timescales derived solely from shell thickness and expansion velocity are prone to substantial uncertainties. 
From the halo mass, we estimate an envelope mass-loss rate of approximately $6 \times 10^{-6}\, M_{\odot}\ \text{yr}^{-1}$ during this pulse. However, this value carries significant uncertainties, primarily due to the poorly constrained duration timescale of the final thermal pulse. This rough estimate broadly aligns with the peak mass-loss rates observed in typical AGB stars. 
Subsequently, the mass-loss rate declined sharply until around 2000 years ago, when the EE and torus formed, signaling the initiation of a new, non-uniform mass-loss episode. In the absence of detailed radiative transfer models, disentangling the mass-loss rates of the EE and torus remains infeasible. Instead, we estimate an average post-AGB nebular mass-loss rate of approximately $7.5 \times 10^{-7}\, M_{\odot}\ \text{yr}^{-1}$.

The outer halo structure represents material shed during the terminal phase of the AGB, while the emergence of a centrally positioned non-spherical torus signals the nebula's evolution toward the PN stage. Current observations show that EDE structures in PPNs can be categorized into two primary types \citep{2018Galax...6...99L}. The first type is a slowly expanding torus with negligible angular momentum and undetectable rotation, as observed around the 
water-fountain source IRAS~16342$-$3814 \citep{2017ApJ...835L..13S} and the 21\,$\mu$m source IRAS~23304+6147 \citep{2025A&A...696A.102S}. The second type is a Keplerian disc with significant angular momentum and distinct rotational motion, such as the circumbinary discs in the Red Rectangle \citep{2016A&A...593A..92B} and around AC~Her \citep{2015A&A...575L...7B}. In IRAS~06530$-$0213, we detect no rotational signature, the expansion velocity is low, and its P-V diagram along the equatorial plane closely resembles that of IRAS~23304+6147, classifying its torus as the first type.

According to \citet{2013A&ARv..21...59I}, the initial torus forms when the companion's gravitational field and tidal forces focus the primary's slow isotropic wind toward the equatorial plane. As the system enters the common-envelope (CE) phase, tidal interactions drive a rapid increase in mass transfer, causing the companion and donor core to spiral inward while releasing orbital energy and angular momentum that ejects the envelope. This interaction occurs on a timescale of only months to years \citep{2022MNRAS.517.3181G}. Material that fails to escape falls back under gravity and reconverges in the equatorial plane, further thickening and stabilizing the torus. This process demonstrates how the CE phase can substantially thicken an early-formed torus, potentially explaining the vertically thickened, barrel-shaped morphology of the observed torus.

Overall, our observations indicate that IRAS~06530$-$0213 experienced its final thermal pulse approximately 6500 years ago, with the pulse duration lasting roughly 650 years. This event produced a spherically symmetric outer halo. Around 4500 years ago, the system entered the  PPN stage, during which the companion's gravitational focusing and wind–Roche-lobe overflow began channeling the AGB star's slow wind into the equatorial plane, gradually assembling an initial torus. 
Thereafter, the system transitioned into the CE phase. Over a timescale of years to decades, the companion and stellar core spiraled inward, driving the global ejection of the envelope. Residual gas unable to reach escape velocity succumbed to gravitational pull, cascading back towards the central region and reconverging in the equatorial plane. This inflow of material further augmented the disk's thickness, ultimately sculpting a barrel-shaped torus.

\subsection{Origin of blue-end narrow CO emission  lines from the PPN halo}\label{co}

\begin{figure*}[ht]
  \centering
  \includegraphics[width=0.49\linewidth]{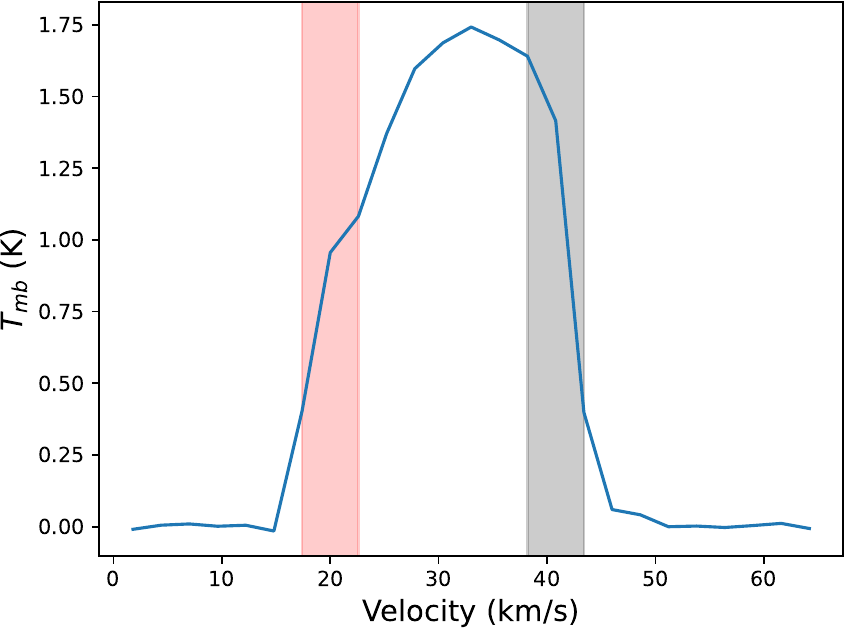}
  \vspace{4em} 
  \includegraphics[width=0.45\linewidth]{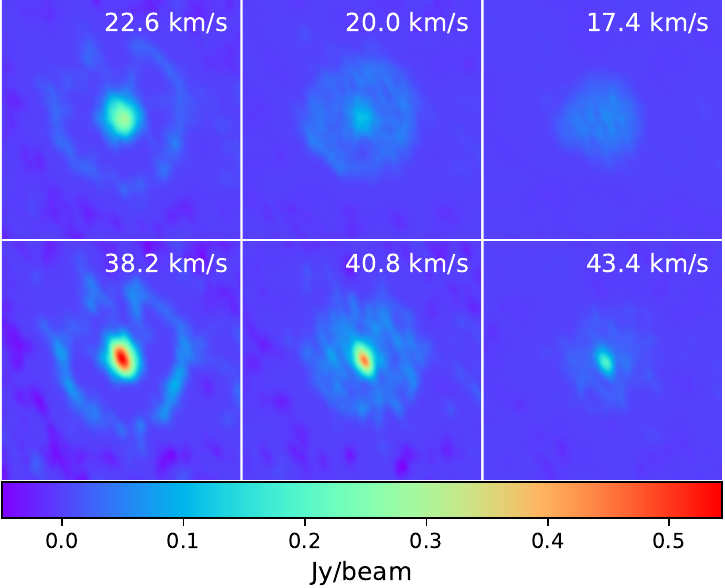}
  \caption{CO ($J=2$--1) emission. The left panel displays the profile derived from averaging flux densities over a $12\arcsec \times 12\arcsec$ area centered at the phase center. The right panel presents the channel maps corresponding to the pink mask (17.4--22.6 $\rm km\,s^{-1}$) and gray mask (38.2--43.4 $\rm km\,s^{-1}$) regions marked in the left panel.}
  \label{co_compare}
\end{figure*}

The single-dish observations targeting IRAS~06530$-$0213 revealed a narrow spectral feature with a width of 2\,km~s$^{-1}$, centered at approximately 20\,km~s$^{-1}$ in the CO ($J$=1--0) transition \citep[see Figure~3 in][]{2024AJ....167...91Q}. This feature was tentatively attributed to contamination by a nearby interstellar cloud within the telescope beam. To investigate this interpretation, we analyze the NOEMA CO ($J$=2--1) line maps in the $V_{\rm LSR}$ ranges of 17.4 to 22.6\,km~s$^{-1}$ and 38.2 to 43.4\,km~s$^{-1}$ (Figure~\ref{co_compare}).
Our results show that due to the obscuration of the central compact structure by the halo in the blue-side channels 
($V_{\rm LSR}\leq20.0$\,km~s$^{-1}$), 
the flux reduction in these channels is more pronounced than in the red-side channels  ($V_{\rm LSR}\geq40.8$\,km~s$^{-1}$) (see the right panel of Figure~\ref{co_compare}). Consequently, the CO profile exhibits a dip around 20\,km~s$^{-1}$. 
Since CO ($J$=1--0) has a lower critical density and excitation temperature than CO ($J$=2--1), it primarily traces the tenuous, cold outer shell where high optical depth induces stronger self-absorption of interior nebular emission. This halo self-absorption produces a dip on the blue side near the systemic velocity in the line profile, yielding the emission on the blue side of the dip (narrow 2\,km~s$^{-1}$ feature) at $\sim$20\,km~s$^{-1}$ that could be misinterpreted as interstellar medium contamination.

The influence of the nebular structure on the IRAS~06530$-$0213 line profile is instructive, suggesting that past analyses based on single-dish data may have been susceptible to misinterpretation or omission of critical information. \citet{2018A&A...618A..91G} presented single-dish CO ($J$=3--2) line observations of 93 PPN and PN using the Atacama Pathfinder EXperiment (APEX) telescope. Notably, the detected CO lines fall into two groups: the first comprises broader lines with velocities matching the source's heliocentric velocity (V$_{\rm hel}$), attributed to the nebula itself; the second, detected mostly in PN, has central velocities inconsistent with the objects' systemic velocities, making interstellar medium (ISM) origin more likely. However, our observations demonstrate that when multiple independent structures formed at different nebular evolutionary stages are observed concurrently, CO emission cannot be attributed to a single velocity component.

\begin{figure*}
\centering
\includegraphics[width=0.8\linewidth]{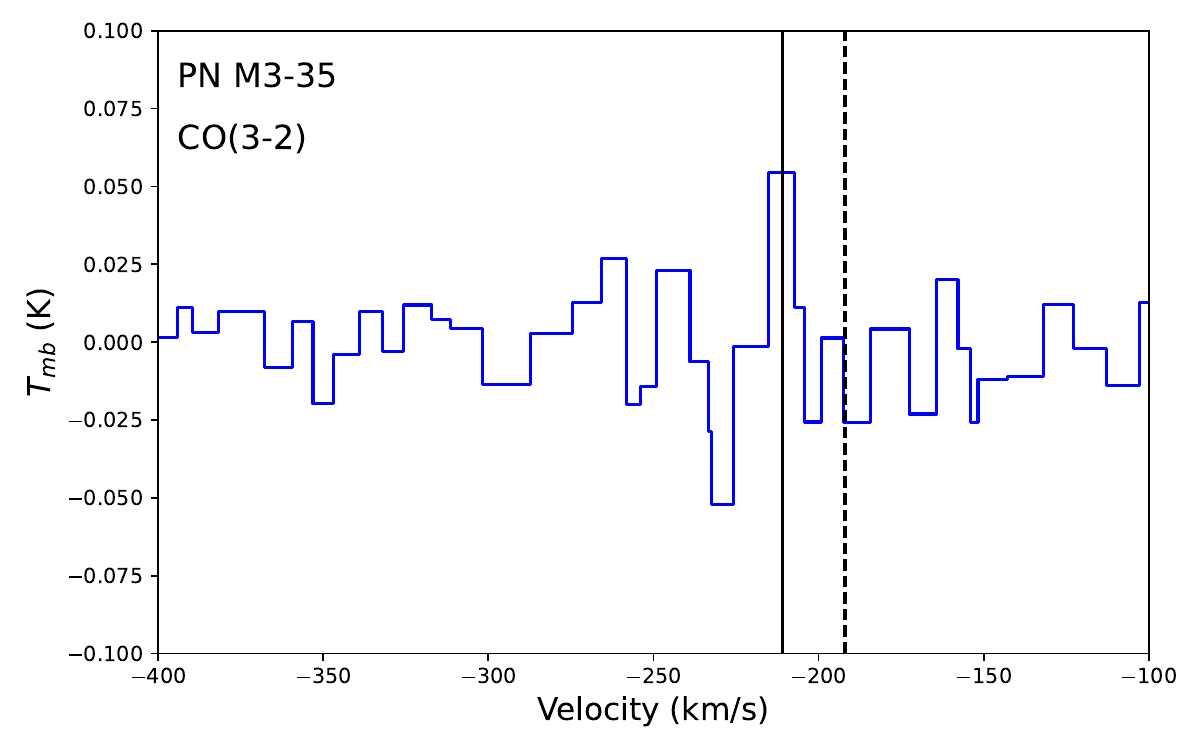}
\caption{CO ($J=3$--2) line in  PN M3-35 \citep[data taken from][]{2018A&A...618A..91G}. The dashed line denotes the heliocentric velocity of  PN M3-35, while the solid line marks the central velocity of the CO $J=3$--2 line.}
\label{m_3-35}
\end{figure*}

By examining Figure~3 of \citet{2018A&A...618A..91G} we
can identify three sources exhibiting CO emission at the blue-end of the heliocentric velocity: PN M3-35, HD 101584, and NGC 6326. Taking PN M3-35 as a representative case, we re-examined its CO ($J$=3--2) line, which was originally ascribed  to interstellar contamination (Figure~\ref{m_3-35}).
PN M3--35 has V$_{\rm hel} = -192.2$\,km~s$^{-1}$; its CO emission is weak and narrow (\textless~5\,km~s$^{-1}$), with the line profile centered at $-210$\,km~s$^{-1}$. Due to its narrow width and offset from the nebula's V$_{\rm hel}$, this feature was originally assigned to the ISM. However, since PN M3-35 has reached the PN stage, central-region CO would mostly have been ionized and undetectable. Conversely, if the progenitor experienced intense thermal pulse mass loss during its AGB or PPN phases, residual CO would persist in the nebula's outer region (analogous to the halo around IRAS~06530$-$0213). This neutral residual CO would manifest as an emission feature on the blue side of the line profile. Thus, we predict that as IRAS~06530$-$0213 evolves into the PN stage, its halo CO emission will similarly appear on the blue side of the systemic velocity.

\section{Conclusion} \label{sec:concl}

We present high-resolution NOEMA interferometric imaging observations of CO and $^{13}$CO ($J$=2--1) emission toward the 21\,$\mu$m PPN IRAS~06530$-$0213, These observations reveal an extended spherical halo and a prominent barrel-shaped torus structure, which collectively highlight the critical evolutionary transition from the AGB  to the PPN phase.
The outer nebular halo is inferred to have originated from the final thermal pulse during the late AGB phase approximately 6500 years ago, while the barrel-shaped torus in the central region might originate from material fallback following a common-envelope episode in the binary system after the source entered the PPN phase around
4500 years ago. The thin halo displays a clumpy and fragmented structure.
The total nebular mass is estimated to be $\sim5.5 \times 10^{-3}\,M_\odot$, with the outer halo exhibiting a mass-loss rate of $\sim6 \times 10^{-6}\,M_\odot$\,yr$^{-1}$ and the central region showing a lower rate of $\sim7.5 \times 10^{-7}\,M_\odot$\,yr$^{-1}$. 
The halo's obscuration of the central nebula induces a depression in the blue wing of the CO line profile, demonstrating that the emission feature at $\sim$20\,km~s$^{-1}$, previously attributed to ISM contamination, originates from the circumstellar halo. 
This interpretation can be extended to several other PPNs that also exhibit blue-shifted CO line emissions.
These results would advance our understanding of mass-loss  processes during the AGB-PPN transition and the role of binary interactions in shaping nebular morphology. Notably, torus-like structures have been consistently detected in 
all 21\,$\mu$m sources observed to date using high-resolution interferometric arrays. This morphological commonality offers a crucial diagnostic clue for constraining the origin of this enigmatic infrared emission band.

\begin{acknowledgments}

We thank the anonymous reviewer for their insightful comments and constructive suggestions, which have greatly improved this manuscript.
We are grateful to 
Ana Lopez Sepulcre and Michael Bremer for their helps in the data reduction. 
The financial supports of this work are from 
the National Natural Science Foundation of China (NSFC, No.\,12473027 and 12333005), the Guangdong Basic and Applied Basic Research Funding (No.\,2024A1515010798), and the Greater Bay Area Branch of the National Astronomical Data Center (No.\,2024B1212080003). 
Sheng-Li Qin acknowledges the financial support from NSFC
(No.\,12033005).
This work is based on observations carried out
with the IRAM NOEMA Interferometer. IRAM is supported by INSU/CNRS (France), MPG (Germany) and IGN (Spain). 

\end{acknowledgments}

\bibliography{sample631}{}
\bibliographystyle{aasjournal}

\newpage

\begin{deluxetable}{lcccccc}
\tablecaption{Observational setup.\label{tab1}}
\tablewidth{0pt}
\tabletypesize{\footnotesize}
\tablehead{
\colhead{Obs. date} & \colhead{Int.Time} & \colhead{Config.} &
\multicolumn{4}{c}{Calibration source} \\
\cline{4-7}
\colhead{} & \colhead{} & \colhead{} &
\colhead{Amplitude} & \colhead{Phase} & \colhead{RF Bandpass} & \colhead{Flux\tablenotemark{a}}
}
\startdata
03-mar-2022 & 3.0h & 12A & J0656-0323 & J0641-0320 & 3C84 & LKHA101 , 2010+723 \\
07-mar-2022 & 3.0h & 12A & J0656-0323 & J0641-0320 & 3C84 & LKHA101 \\
11-oct-2021 & 2.2h & 10D & 0723-008   & J0656-0323 & 3C84 & LKHA101 \\
14-oct-2021 & 1.5h & 10D & 0723-008   & J0656-0323 & 3C84 & LKHA101 \\
\enddata
\tablenotetext{a}{
For LKHA 101, the flux density is 0.56 Jy at 231 GHz, with a spectral index of 0.96. For 2010+723, the flux density is 0.27 Jy at 231 GHz, with a spectral index of $-0.68$.}
\end{deluxetable}

\begin{deluxetable}{lcccccc}
\tablecaption{CO and its isotopologue lines \label{tab:example}}
\tablehead{
  \colhead{Molecule ID} & \colhead{Transition} & \colhead{$E_{\rm u}$} & \colhead{Frequency} 
    & \colhead{$I_{\mathrm{peak}}$} & \colhead{$\int I\,dv$} & \colhead{rms} \\
  & & \colhead{(K)} & \colhead{(MHz)} 
    & \colhead{(K)} & \colhead{($\mathrm{K\,km\,s}^{-1}$)} & \colhead{(K)}
}
\startdata
CO         & $J=2\!-\!1$      & 16.59 & 230538.00 & 22.13 & 165.30 & 0.18 \\
$^{13}$CO  & $J=2\!-\!1$      & 15.68 & 220398.67 &  0.85 &   6.75 & 0.17 \\
C$^{18}$O  & $J=2\!-\!1$      & 15.81 & 219560.35 &  0.20 &   1.39 & 0.17 \\
\enddata
\tablecomments{ Within the 210--230 GHz frequency range, the conversion factor from Jy\,beam$^{-1}$ to K spans 77--86 K/(Jy\,beam$^{-1}$) for the combined dataset.}
\label{line}
\end{deluxetable}

\newpage

\end{document}